\documentstyle[prc,preprint,tighten,aps]{revtex} 
\begin {document}
\draft

\title{Coulomb-Sturmian separable expansion approach:
Three-body Faddeev calculations for Coulomb-like interactions
\footnote{to appear in Physical Review C, Volume 54, 
Issue 1, 1 July 1996}}

\author{Z. Papp \\ 
 Institute of Nuclear Research of the Hungarian Academy
of Sciences, \\
Bem t\'er 18/c, P.O. Box 51, H--4001 Debrecen, Hungary }
\author{ W. Plessas \\
 Institute for Theoretical Physics, University of Graz,\\
Universit\"atsplatz 5, A-8010 Graz, Austria} 

\date{\today}

\maketitle
\begin{abstract}
\noindent

We demonstrate the feasibility and efficiency of the 
Coulomb-Sturmian separable
expansion method for generating accurate solutions of the Faddeev
equations. Results obtained with this method are reported for several
benchmark cases of bosonic and fermionic three-body systems. Correct
bound-state results in agreement with the ones established in the 
literature are
achieved for short-range interactions. We outline the formalism for 
the treatment of three-body Coulomb systems and present a bound-state
calculation for a three-boson system interacting via 
Coulomb plus short-range forces. The corresponding result is in good
agreement with the answer from a recent stochastic-variational-method
calculation.

\end{abstract}
\vspace{0.5cm}
\pacs{PACS number(s): 21.45.+v, 03.65.6e, 02.30.Rz, 02.60.Nm}

\narrowtext

\section{Introduction}

Separable expansion schemes have always been extremely useful in solving
few-body problems. For example, in the three-body systems, at any time
starting from the 60's, important new results have been
achieved by solving the
Faddeev equations with some sort of separable representation of the
two-body subsystems (for a review see, e.g., ref. \cite{plessas1}). 
In more recent times, above all the separable expansion method proposed
by Ernst, Shakin, and Thaler \cite{EST} (the so-called EST method) has
proven very useful. Indeed, the first three-nucleon scattering
results with such realistic meson-exchange N-N interactions like the
Paris potential were achieved along this method \cite{koike},
and they were later on confirmed by a direct solution of the Faddeev
equations \cite{cornelius}.

Even nowadays when three-body Faddeev equations can directly be solved
on supercomputers (for a recent review of the state of the art see ref.
\cite{glockle}), separable expansion schemes have their relevance.
Not only does an accurate separable representation of the input dynamics
allow to save much computer time in arriving at standard results,
separable expansion methods may also help a lot in obtaining solutions to
hitherto unsolved problems. In this respect we may mention the solution
of the three-nucleon scattering problem with realistic N-N interactions
and Coulomb forces at any energy; so far only limited solutions below
 or above  breakup threshold have been
obtained (see, e.g., refs. \cite{berthold} and \cite{alt}).

In this paper we deal with a separable expansion method that is
well-adapted to treating few-body problems including
long-range forces. Its
essence lies in the expansion of the potential operator $v^s$ of the
short-range part of any interaction with the use of 
Coulomb-Sturmian (CS) functions
\cite{papp1}. If the full potential also contains a Coulomb interaction
$v^C$, this is kept in the Green's operator. Thereby all difficulties
associated with a (separable) expansion of the Coulomb potential are
avoided, while at the same time correct asymptotic properties of all
quantities are guaranteed. Still the advantages of the separable
expansion can be exploited in solving the two- and three-body integral
equations.

The CS separable expansion method has been extensively tested before in
two-body problems. Not only bound and resonant states with a variety of
short-range plus Coulomb potentials \cite{papp1} were investigated but
also scattering solutions were obtained \cite{papp2}. 
In these works also convergence
studies were performed and subsequently extended to the 
multichannel Coulomb
problem \cite{pzews}. Computer codes for the CS separable expansion of any
local or non-local two-body interaction under the presence of 
Coulomb-like potentials were published in ref. \cite{cpc}. 
In Ch. II below we shall recall some of the most important 
formulae for the two-body problem.

The principal advantage of the CS separable expansion is the fact that
the matrix elements of the Coulomb Green's operator can be calculated
analytically in the two-body system. Besides the separable
representation of the short-range part of the full interaction this
turns out to be an essential requirement for an efficient and accurate
solution of the three-body system. For the latter we may thus follow 
the integral-equation approach and thereby guarantee for the
implementation of the appropriate (Coulomb-like) asymptotics. In Ch. III
below we shall demonstrate how the matrix elements of the three-body
Coulomb Green's operator can be calculated in a reliable way.

We prove the efficiency of our method through the solution of the Faddeev
equations for three-body bound states interacting via various
short-range forces and a case with additional Coulomb interaction among
all three particles. We adhere to problems for which benchmark results
from other methods have been obtained already. It is found that in all
cases excellent agreement is achieved. The method therefore appears
promising as an efficient tool for solving three-body systems, as it
can be adapted to more general cases (including three-body forces) and
extended to scattering under the presence of long-range interactions.

\section{Coulomb-Sturmian separable expansion}

We give a short account of the formalism of the CS separable expansion
in the two-body system with short-range plus Coulomb interactions. 
Here we specify our notation and provide the formulae that are 
later on needed
in the solution of the three-body problem.

\subsection{Basis functions}

The short-range potential operator $v^s$ in some angular-momentum state $l$
will be expanded on the basis of Coulomb-Sturmian functions
\begin{eqnarray}
\langle r|nl\rangle=
\left[\frac{n!}{(n+2l+1)!}\right]^{1/2} (2br)^{l+1} e^{-br} 
L_n^{2l+1} (2br), \nonumber \\
\ \ \ \ \ \ \ \ \ \ \ \ \ \ \ \ \  (n=0,1,2,...),
\label{basis}
\end{eqnarray}
which are the solutions of the Sturm-Liouville problem of the hydrogenic
system \cite{rotenberg}. Here, $L_n^{2l+1}$ represent the Laguerre
polynomials and $b$ relates
to the energy in the Sturm-Liouville equation. 
We take $b$ as a fixed parameter thus working with energy-independent
CS functions. They form a complete set
\begin{equation}
{\bf{1}}=
\lim_{N\to\infty} \sum_{n=0}^N  |\widetilde{nl}\rangle 
\langle nl| = \lim_{N\to\infty} {\bf{1}}_N,
\label{unity}
\end{equation}
where
\begin{equation}
\langle r|\widetilde{nl}\rangle=\frac{1}{r} \langle r|nl\rangle .
\end{equation}

With the N-th order unit operator ${\bf 1}_N$ in Eq. (\ref{unity}) we can
now expand the short-range potential operator in the form
\begin{equation}
v^s_l=\lim_{N\to\infty} {\bf 1}_N  v^s_l {\bf 1}_N 
=\lim_{N\to\infty} \sum_{n,n'=0}^N   |\widetilde{nl}\rangle 
\langle nl| v^s_l | n'l \rangle \langle \widetilde{n'l} | .
\label{expa} 
\end{equation}
If $N$ remains finite, we end up with a rank-N separable approximation. 
As a consequence the two-body problem can then be solved by 
algebraic methods \cite{plessas1}.

\subsection{Short-range plus Coulomb interactions}

Let us now assume a two-potential case of short-range plus Coulomb-like
interactions
\begin{equation}
v= v^s + v^C 
\end{equation}
and consider the homogeneous Lippmann-Schwinger equation for the bound state
$|\psi_l\rangle$ in some partial wave $l$
\begin{equation}
|\psi_l\rangle=g_l^C(E) v^s |\psi_l\rangle.
\label{LS}
\end{equation}
Here $g_l^C(E)$ is the two-body Coulomb Green's operator
\begin{equation}
g_l^C(E)=(E-h_l^0-v^C)^{-1}
\end{equation}
with the free Hamiltonian denoted by $h_l^0$. 
Using the expansion (\ref{expa}) in Eq. (\ref{LS}) one arrives at a 
linear system of homogeneous quations for the wave-function coefficients
$\underline{A}_{ln}= \langle \widetilde{nl} | \psi_l \rangle$:
\begin{equation}
 [(\underline{g}_l^C(E))^{-1} - \underline{v}^s_l ] 
 \underline{A}_{l} =0.
\label{eq18a}
\end{equation}
It has a unique
solution if and only if
\begin{equation}
\det [(\underline{g}_l^C(E))^{-1} - \underline{v}^s_l ] =0.
\end{equation}
The matrices involved are made up from the elements
\begin{equation}
\underline{g}_{l n n'}^C(E) = \langle \widetilde{nl} | g_l^C (E) | 
 \widetilde{n'l} \rangle 
\end{equation}
and
\begin{equation}
\underline{v}^s_{l n n'} =  \langle {nl} |v^s_l |  {n'l} \rangle .
\end{equation}
While the latter matrix elements may be evaluated (numerically) for any
given short-range potential either in configuration or in momentum space,
the matrix elements of the Coulomb Green's operator between CS states can
be calculated analytically \cite{papp1}; the corresponding computer code
is available from ref. \cite{cpc}. This fact then 
also allows to calculate the matrix elements of the 
full Green's operator  in the whole complex plane,
\begin{equation}
\underline{g}_l(E) =
 ( (\underline{g}_l^C(E))^{-1}-\underline{v}^s_l)^{-1},
\label{2bgreen}
\end{equation}
what will be needed later on in the solution of the three-body problem 
with charged particles.

After solving Eq. (\ref{eq18a}) for the coefficients 
$\underline{A}_{l n}$ the bound state $|\psi_l\rangle$ can be expressed 
as
\begin{equation}
|\psi_l\rangle= \sum_{n=0}^N \underline{B}_{l n} \  g_l^C(E) 
 |\widetilde{nl} \rangle,
\label{11a}
\end{equation}
where the new coefficients result from the matrix multiplication
$\underline{B}_{l}=\underline{v}^s_{l}\underline{A}_{l}$. We note 
that expression (\ref{11a}) is distinct from a usual expansion of the
state $|\psi_l\rangle$ with certain test functions. The explicit
occurence of the Coulomb Green's operator always ensures the correct
asymptotic behaviour \cite{papp2}. This is an immediate consequence
of the fact that only the short-range potential 
(but not the wave function) is expanded.

\section{Solution of the three-body bound-state problem}

We now extend the CS basis to the three-body system and demonstrate the
solution of the Faddeev equations for bound states of three particles 
with any short-range interactions and under the presence of Coulomb
forces.

\subsection{Short-range interactions}

The integral equations  for the three Faddeev components $ \Psi_\alpha $
of the bound-state wave function $ \Psi $ read:
\begin{equation}
|\Psi_\alpha \rangle = G_\alpha (E) [ v^s_\alpha | \Psi_\beta \rangle
 + v^s_\alpha | \Psi_\gamma \rangle ]
\end{equation}
with $\alpha$,$\beta$,$\gamma$ a cyclic permutation. 
Here the channel Green's operators
are defined by
\begin{equation}
G_\alpha (E) = (E - H^0 - v_\alpha^s )^{-1},
\end{equation}
where $H^0$ is the free three-particle Hamiltonian and $v_\alpha^s$ the 
short-range interaction of the pair ($\beta$,$\gamma$). In the
angular-momentum representation (omitting the explicit spin and isospin
dependence from our notation) we define the CS basis for the expansion of
the short-range interactions in the three-particle system as
\begin{equation}
| n \nu l \lambda \rangle_\alpha = | n l \rangle_\alpha \otimes 
| \nu \lambda \rangle_\alpha , \ \ \ \ (n,\nu=0,1,2,...),
\label{cs3}
\end{equation}
with the CS states from Eq. (\ref{basis}). 
Here $l$ and $\lambda$ denote the angular
momenta of the two-body pair ($\beta$,$\gamma$) and of the third particle
$\alpha$ relative to the centre of mass of this pair, respectively. In
the three-particle Hilbert space we have
\begin{equation}
{\bf 1} =\lim_{N\to\infty} \sum_{n,\nu=0}^N
| \widetilde{n \nu l \lambda} \rangle_\alpha \ 
\mbox{}_\alpha\langle  {n \nu l \lambda} | = 
\lim_{N\to\infty} {\bf 1}_{N,\alpha}
\end{equation}
where the configuration-space representation in terms of Jacobi coordinates
$\xi_\alpha$ and $\eta_\alpha$ reads:
\begin{equation}
\langle \xi_\alpha \eta_\alpha |\widetilde{ n \nu l \lambda }
\rangle_\alpha=
\frac{1}{\xi_\alpha \eta_\alpha}
 \langle \xi_\alpha \eta_\alpha |{ n \nu l \lambda }\rangle_\alpha .
\end{equation}
After the CS expansion of the potentials $v^s_\alpha$, $v^s_\beta$, and 
$v^s_\gamma$ in the three-particle space, the Faddeev equations 
can be rewritten as
\begin{equation}
|\Psi_\alpha \rangle = G_\alpha (E) 
\left[{\bf 1}_{N,\alpha} v^s_\alpha {\bf 1}_{N,\beta} |\Psi_\beta \rangle +
{\bf 1}_{N,\alpha} v^s_\alpha {\bf 1}_{N,\gamma} 
|\Psi_\gamma \rangle\right].
\label{sepfe}
\end{equation}
By applying the CS states $\mbox{}_\alpha \langle 
\widetilde{ n \nu l \lambda }| $ from the left, 
Eqs. (\ref{sepfe}) turn into a
linear system of homogeneous equations for the coefficients of the
Faddeev components $\underline{A}_{l_\alpha \lambda_\alpha n \nu }=
\mbox{}_\alpha \langle \widetilde{ n \nu l \lambda}|\Psi_\alpha
\rangle$:
\begin{equation}
 [(\underline{G}(E))^{-1} - \underline{v} ] \underline{A} =0.
\label{fep1}
\end{equation}
A unique solution threreof exists if and only if 
\begin{equation}
\det [(\underline{G}(E))^{-1} - \underline{v} ] =0.
\label{blockst}
\end{equation}
The matrices $\underline{G}(E)$ and $\underline{v}$ have a block structure
and the matrix elements are given by 
\begin{equation}
\underline{v}_{l_\alpha \lambda_\alpha n \nu  ,
{l'}_\beta {\lambda'}_\beta n' \nu' } =
(1-\delta_{\alpha \beta})\  
\mbox{}_\alpha\langle n \nu l \lambda |v^s_\alpha| 
 n' \nu' l' {\lambda}' \rangle_\beta  
\label{vab}
\end{equation}
and
\begin{equation}
\underline{G}_{l_\alpha \lambda_\alpha n \nu ,
{l'}_\alpha {\lambda'}_\alpha n' \nu' } (E) =
\delta_{\alpha \beta}\ 
\mbox{}_\alpha\langle \widetilde{ n \nu l \lambda }|G_\alpha(E)| 
\widetilde{ n' \nu' {l'} {\lambda'} }\rangle_\alpha,
\label{G}
\end{equation}
respectively.
Notice that the matrix elements of the Green's operator are needed only between 
the same partition $\alpha$ whereas the matrix elements of the 
potentials occur only
between different partitions $\alpha$ and $\beta$. 
The latter may again be evaluated numerically
either in configuration or momentum space. We have adopted the 
configuration-space 
version of the Balian-Br\'ezin method \cite{bb}.

For the calculation of the matrix elements of the Green's operator
in Eq. (\ref{G}) we proceed in the following way. We split the three-particle
free Hamiltonian into 
\begin{equation}
H^0=h^0_{\xi_\alpha}+h^0_{\eta_\alpha},
\end{equation}
i.e. the free motions in the two Jacobi coordinates. Then we define the
two-body Hamiltonian $h_{\xi_\alpha}$ as
\begin{equation}
h_{\xi_\alpha}=h^0_{\xi_\alpha}+v^s_\alpha.
\end{equation}
Since the commutator of the Hamiltonians $h_{\xi_\alpha}$ and
$h^0_{\eta_\alpha}$ vanishes,
\begin{equation}
\left[h_{\xi_\alpha},h^0_{\eta_\alpha}\right]=0,
\end{equation}
we may apply the convolution theorem by Bianchi and Favella
\cite{bianchi}
\widetext
\begin{equation}
G_\alpha (E)=(E-h_{\xi_\alpha}-h^0_{\eta_\alpha})^{-1} 
=\frac{1}{2 \pi i} \oint_C d\epsilon (E-\epsilon - h_{\xi_\alpha})^{-1}
 (\epsilon - h^0_{\eta_\alpha})^{-1}.
\label{contourint}
\end{equation}
Here the contour $C$ encircles the spectrum of $h^0_{\eta_\alpha}$
without penetrating into the spectrum of $h_{\xi_\alpha}$ 
(cf. Fig. \ref{fig1}). 
We note that in this integral the roles of
$h^0_{\eta_\alpha}$ and $h_{\xi_\alpha}$ may also be interchanged.

After sandwiching the above Green's operator between the CS states, the
integral in Eq. (\ref{contourint}) appears in the form
\begin{equation}
\underline{G}_{l_\alpha \lambda_\alpha n \nu ,
l_\alpha' \lambda_\alpha' n' \nu' } (E)=
\frac{1}{2 \pi i} \oint_C d\epsilon \ 
\mbox{}_\alpha\langle \widetilde{ n  l}| 
(E-\epsilon - h_{\xi_\alpha})^{-1}
|\widetilde{ n' {l'}}\rangle_\alpha
\ \mbox{}_\alpha\langle \widetilde{ \nu  \lambda }|
(\epsilon - h^0_{\eta_\alpha})^{-1}	
| \widetilde{\nu' {\lambda'} }\rangle_\alpha, 
\end{equation}
\narrowtext
\noindent
where the separate matrix elements occurring in the integrand are known
from the two-particle case of the previous chapter 
(cf. Eq. (\ref{2bgreen})).

After solving Eq. (\ref{fep1}) for the coefficients 
$\underline{A}_{l_\alpha \lambda_\alpha n \nu }$ the
Faddeev components can be expressed as 
\begin{equation}
|\Psi_\alpha\rangle= \sum_{n,\nu=0}^N \underline{B}_{\alpha n \nu} 
\  G_\alpha (E) 
 |\widetilde{n \nu l \lambda} \rangle_\alpha,
\label{fewf}
\end{equation}
where the new coefficients again result from the matrix 
multiplication $\underline{B}=\underline{v}\;\underline{A}$.
As before we hint to the advantage that the representation
(\ref{fewf}) of $|\Psi_\alpha\rangle$ guarantees for the correct
asymptotic behaviour of the Faddeev component, due to the explicit 
occurrence of the Green's operator. This is of particular importance in
the Coulomb case below.

\subsection{Coulomb-like interactions}

In this section we extend the formulation of the three-body problem to
the case of long-range interactions. We assume the subsystem interaction
to be a sum of short-range plus (repulsive) Coulomb interactions
\begin{equation}
v^\alpha=v^s_\alpha+v^C_\alpha,
\end{equation}
and adhere to the Faddeev equations in the form as modified by Noble
\cite{noble}:
\begin{equation}
|\Psi_\alpha \rangle = G^C_\alpha (E) [ v^s_\alpha | \Psi_\beta \rangle
 + v^s_\alpha | \Psi_\gamma \rangle ]
\label{coufewf}
\end{equation}
with the Coulomb-like Green's operator
\begin{equation}
G^C_\alpha (E) = (E - H^0 - v_\alpha^s - v_\alpha^C - v_\beta^C -
v_\gamma^C)^{-1}.
\label{3CG}
\end{equation}
Herein all long-range interactions for all subsystems are collected, in
complete analogy to the two-body case (cf. Eq. (\ref{LS})).

As in the previous section the short-range potentials are expanded on the CS
basis leading to an equation similar to (\ref{sepfe}). A unique solution
thereof exists if and only if 
\begin{equation}
\det [(\underline{G}^C (E))^{-1} - \underline{v} ] =0,
\end{equation}
where the matrices have the same block structure as before in
Eq. (\ref{blockst}). The important point is that $\underline{v}$ contains
only matrix elements of the short-range interactions; in fact they are
completely equivalent to Eq. (\ref{vab}).

For the calculation of the matrix elements of the Coulomb-like Green's
operator we proceed along the lines of the two-potential 
formalism \cite{bencze}. First we rewrite it using the
resolvent equation
\begin{equation}
G^C_\alpha (E) = \widetilde{G}^C_\alpha (E) + \widetilde{G}^C_\alpha (E)
 (v_\beta^C + v_\gamma^C - u_\alpha^C )   G^C_\alpha (E)
\label{resgc}
\end{equation}
with $\widetilde{G}^C$ defined by
\begin{equation}
\widetilde{G}^C_\alpha (E)=( E-H^0-v_\alpha^s-v_\alpha^C-u_\alpha^C )^{-1}.
\label{3gc}
\end{equation}
Here we have introduced the auxiliary potential $u_\alpha^C$, which
is required to have the asymptotic form
\begin{equation}
u_\alpha^C \sim \frac{Z_\alpha
(Z_\beta+Z_\gamma) }{\eta_\alpha}
\end{equation}
as ${\eta_\alpha \to \infty}$.
It may be viewed as the effective Coulomb potential between the centre
of mass of the subsystem $\alpha$ (with charge $Z_\beta+Z_\gamma$)
and the third particle (with charge $Z_\alpha$). The important role of
the potential $u_\alpha^C$ is that asymptotically 
it compensates the Coulomb tail
of the long-range potentials $v_\beta^C + v_\gamma^C$ in Eq. (\ref{resgc}). 
Thus the combination $U=v_\beta^C + v_\gamma^C - u_\alpha^C$
can be subject to a separable expansion and effectively
be treated as a short-range potential.
With the help of the formal solution of Eq. (\ref{resgc})
we may now express the inverse matrix $(\underline{G}^C_\alpha (E))^{-1}$
as 
\begin{equation}
(\underline{G}^C_\alpha (E))^{-1}=(\underline{\widetilde{G}}^C_\alpha (E))^{-1}
-\underline{U},
\end{equation}
where $\underline{U}$ is constructed from the matrix elements
\begin{equation}
\underline{U}_{l_\alpha \lambda_\alpha n \nu ,
l'_\alpha {\lambda}'_\alpha  n' \nu' } =
\mbox{}_\alpha\langle n \nu l \lambda | (v_\beta^C + v_\gamma^C -
u_\alpha^C)
| n' \nu' l' {\lambda}' \rangle_\alpha 
\end{equation}
resulting from the CS expansion. The matrix elements of the Green's operator 
$\widetilde{G}^C_\alpha(E)$ can again be calculated by the contour
integral as before. We are then left with the integral
\widetext
\begin{equation}
\underline{\widetilde{G}}^C_{l_\alpha 
\lambda_\alpha n \nu ,l_\alpha' \lambda_\alpha' n' \nu' } (E)=
\frac{1}{2 \pi i} \oint_C d\epsilon \ 
\mbox{}_\alpha \langle \widetilde{ n  l }| 
(E-\epsilon - h_{\xi_\alpha} )^{-1}
|\widetilde{ n' {l'}  }\rangle_\alpha \ 
\mbox{}_\alpha\langle \widetilde{ \nu  \lambda }|(\epsilon - 
h^C_{\eta_\alpha} )^{-1}	
| \widetilde{\nu' {\lambda'}}\rangle_\alpha ,
\end{equation}
\narrowtext
\noindent
where $h^C_{\eta_\alpha}=h^0_{\eta_\alpha}-u^C_\alpha$ and 
$h_{\xi_\alpha}=h^0_{\xi_\alpha} - v^s_\alpha - v^C_\alpha$.

The calculation of the Faddeev amplitude $|\Psi_\alpha \rangle$ in
Eq. (\ref{coufewf}) is completely analogous to the short-range case
of the previous section. Only in Eq. (\ref{fewf}) the Green's
operator $G_\alpha (E)$ must now be replaced by the Coulomb Green's
operator $G^C_\alpha (E)$ of Eq. (\ref{3CG}).

\section{Tests of the method}

In this chapter we demonstrate the performance of the method in
calculations of various three-body bound states. We have selected 
cases for which benchmark results are already available in the
literature. The comparisons will prove the efficiency of our method,
especially in the situation when Coulomb forces are present.

\subsection{Illustration of the convergence of the CS expansion}

Before presenting the final results, let us demonstrate the convergence
of the results for the three-body bound-state energies. For this purpose
 we take the example of the Ali-Bodmer (AB) potential \cite{ali} between 
$\alpha$-particles of mass $M$
\begin{equation}
v^s (r) = 500 \exp (-(0.7 r)^2) - 130 \exp (-(0.475 r)^2)
\label{alibodmer}
\end{equation}
without and
\begin{equation}
 v (r) = v^s (r) + 4 \mbox{e}^2/r
\label{alibodmerC}
\end{equation}
with Coulomb interaction. We use units such that
$\hbar^2/M=10.36675\ \mbox{MeV}\ \mbox{fm}^2$
and $\mbox{e}^2=1.44\ \mbox{MeV}\ \mbox{fm}$.

Evidently the quality of the results will depend on the number of terms
employed in the separable expansion of the (short-range) potential.
We quote the values of the binding energies from calculations without
(Table \ref{tab}) and with (Table \ref{tabc}) Coulomb forces, taking into 
account different numbers of channels $n_{ch}=1,2,3,4$ (corresponding 
to angular momentum states up to $l=\lambda=0$, $l=\lambda=2$, $l=\lambda=4$,
and $l=\lambda=6$ employed).
In all cases it is observed that convergence up to $5$ significant
digits is comfortably achieved with $N=20$ terms applied for $n$ and $\nu$
in the separable expansion. Remarkably the speed of the convergence
is everywhere similar, irrespective of how many angular-momentum
channels are included and whether or not Coulomb forces are present. We note
especially for the Coulomb case that the satisfactory convergence stems
from reliable separable expansions of the potentials $v^s$ and $U$, which
- from the point of view of scattering theory - are both short-range
potentials; the fall-off of $U$ is much slower than of $v^s$, though.

In principle, the convergence may also depend on the (range) parameter
$b$ of the Coulomb-Sturmian functions. We found, however, that the
dependence is weak in a relatively large interval of possible choices,
just as it was established in the two-body case \cite{papp1,papp2}.

The convergence is practically of the same quality in the case of the other
potentials considered below.

\subsection{Results for various three-body bound states}

We now present our converged results for the Malfliet-Tjon
(MT) \cite{malflet} and AB \cite{ali} potentials and compare them
to other benchmark calculations. For the MT potential
\begin{equation}
v^s (r) =  V_r \exp (- 3.11 r) /r -V_a \exp (- 1.55 r) /r
\label{mt5}
\end{equation}
between two nucleons of mass m we consider two cases \\
MTVa: $V_r=1458.0470$ $V_a=578.0890$ \\
MTVb: $V_r=1438.4812$ $V_a=570.3316$ \\
(here we use units such that
$\hbar^2/m=41.47\ \mbox{MeV}\ \mbox{fm}^2$).
For MTVa
we may compare to the results of the Los Alamos-Iowa \cite{la1}
and Groningen \cite{gr1} groups, both of which are obtained from
a direct solution of the Faddeev equations in configuration space,
and in addition to results from an ATMS \cite{atms} calculation,
a Green's function Monte-Carlo (GFMC) calculation \cite{gfmc},
an integro-differential-equation approach (IDEA) \cite{idea}, and
a stochastic variational method (SVM) \cite{svm}. From Table 
\ref{tmt5} it is evident that our method provides very accurate
predictions for binding energies
 in all cases. This is true for the channel-by-channel
comparison with the  Los Alamos-Iowa calculation and likewise for
the comparison with the best results from the other works. With respect
to the best results quoted in the lower rows of Table \ref{tmt5} we 
note that for the corresponding calculations the number of
angular-momentum channels employed is either not definitely known or
not specified in  a scheme like ours.

In order to demonstrate that in addition to binding energies our method
also provides accurate three-body wave functions, we calculated the
root-mean-square (rms) radius $<r^2>^{1/2}$. Corresponding results are
given in Table \ref{tmt5b} for the MTVb potential, in which case
we can compare to the calculations of the Los Alamos-Iowa group 
\cite{la1}. For both the binding energy and the rms radius the 
channel-by-channel comparison indicates perfect agreement.

In case of the MTI-III potential for a system of three fermions, acting
in singlet and triplet states, we may compare to the 2-channels
calculation of the Los Alamos-Iowa group. For the MTI-III potential as
parametrized in ref. \cite{la2}, we obtain a binding energy (converged
result) of $E=8.5358$ in comparison to $E=8.536$ calculated by the 
Los Alamos-Iowa collaboration.

Finally we come to the comparison of the results for the binding energy
of a system of three bosons interacting via the AB and AB
plus Coulomb potential (Table \ref{tabcomp}).
We may compare our 4-channels result to calculations with the ATMS
method \cite{atms} (uncharged case only) and the SVM \cite{svm,privat}.
We do not know of any Faddeev results in this case. 
Again we show predictions for the binding energies and rms radii.
All the values quoted in
Table \ref{tabcomp} show a convincing agreement of our results with the
ones from the other approaches. We specially stress the agreement of the
result for the case of rigorously including the Coulomb interaction with
the rather reliable answer from the SVM.

\section{Conclusion}

We have suggested a separable expansion scheme, relying on 
Coulomb-Sturmian basis functions, for solving the three-body problem.
The method is especially suited to the case when Coulomb-like
interactions are present in one or all subsystems. It allows to solve
the three-body integral equations by expanding only the short-range
part of the interaction in a separable form while keeping the effect of
the long-range part in an exact manner via a proper integral
representation of the three-body Coulomb Green's operator. As a
consequence the method has good convergence properties and can 
practically be made arbitrarily accurate by employing an increasing number
of terms in the separable expansion. The usage of the Coulomb-Sturmian
basis is essential to allow for the accurate evaluation of the matrix
elements of the Coulomb Green's operator.

Beyond the studies of the method in systems with two-body asymptotics
conducted before
\cite{papp1,papp2,cpc,pzews}, we have now demonstrated its convergence 
properties and efficiency in
(benchmark) calculations of the three-body bound-state problem without
and with Coulomb interactions.
In both cases the solution of the Faddeev equations show 
a rapid convergence, and, 
whenever a comparison is possible to existing results in
the literature, correct predictions for the binding energies 
and wave functions are achieved.

The method is capable of treating any kind of short-range interactions,
even in the case when Coulomb-like forces are present. The solution of
the three-body bound-state problem was carried out here. However, the
method is also applicable for scattering problems. In this regard it has
been proven useful already in the two-body case \cite{papp2,cpc,pzews}.
To solve the corresponding problem for a (charged) three-body system 
with the 
Faddeev equations some technical details in connection with the evaluation
of the then occurring matrix elements still need to be worked out.

\acknowledgments

The authors are grateful to K. Varga for providing results from his SVM
calculation for comparison. The work of Z. P. was supported by OTKA under
contracts F4305 and T17298. This work received support also from
the Austrian-Hungarian Scientific-Technical Cooperation within 
project A23.

\narrowtext

\begin{table}
\caption{
Convergence of the binding energy of a three-boson system interacting
via the Ali-Bodmer potential, Eq. (\ref{alibodmer}), with increasing
basis for the separable expansion. $N$ denotes the maximum number of
basis states employed for $n$ and $\nu$ in Eq. (\ref{sepfe}).}
\label{tab}
\begin{tabular}{rcccc}
$N$ &  \multicolumn{4}{c}{\mbox{Number of channels $n_{ch}$} }  \\
    & 1      & 2       &       3 &    4   \\
\hline 
12 & 4.13899 & 5.11756 & 5.17699 & 5.17888 \\ 
13 & 4.14092 & 5.11911 & 5.17862 & 5.18057 \\ 
14 & 4.14044 & 5.11897 & 5.17846 & 5.18043 \\ 
15 & 4.14046 & 5.11898 & 5.17850 & 5.18047 \\ 
16 & 4.14065 & 5.11917 & 5.17871 & 5.18069 \\ 
17 & 4.14069 & 5.11926 & 5.17880 & 5.18079 \\ 
18 & 4.14069 & 5.11926 & 5.17881 & 5.18080 \\ 
19 & 4.14069 & 5.11926 & 5.17881 & 5.18080 \\ 
20 & 4.14070 & 5.11927 & 5.17882 & 5.18081 \\ 
21 & 4.14071 & 5.11928 & 5.17884 & 5.18083 \\ 
22 & 4.14071 & 5.11929 & 5.17884 & 5.18083 \\ 
23 & 4.14071 & 5.11929 & 5.17884 & 5.18083 \\ 
24 & 4.14071 & 5.11929 & 5.17884 & 5.18083 \\ 

\end{tabular}
\end{table} 

\begin{table}
\caption{
Same as in Table \ref{tab} for the Ali-Bodmer plus Coulomb
potential, Eq. (\ref{alibodmerC}).}
\label{tabc}
\begin{tabular}{rcccc}
$N$ &  \multicolumn{4}{c}{\mbox{Number of channels $n_{ch}$} }  \\
    & 1      & 2       &       3 &    4   \\
\hline 
12 & 1.90151 & 2.81629 & 2.86703 & 2.86839 \\
13 & 1.90473 & 2.81833 & 2.86912 & 2.87054 \\ 
14 & 1.90368 & 2.81796 & 2.86871 & 2.87014 \\ 
15 & 1.90373 & 2.81797 & 2.86875 & 2.87019 \\ 
16 & 1.90404 & 2.81819 & 2.86899 & 2.87044 \\ 
17 & 1.90397 & 2.81824 & 2.86904 & 2.87049 \\ 
18 & 1.90401 & 2.81824 & 2.86905 & 2.87050 \\ 
19 & 1.90400 & 2.81824 & 2.86905 & 2.87050 \\ 
20 & 1.90401 & 2.81825 & 2.86906 & 2.87051 \\ 
21 & 1.90402 & 2.81826 & 2.86907 & 2.87053 \\ 
23 & 1.90402 & 2.81827 & 2.86908 & 2.87053 \\ 
23 & 1.90402 & 2.81827 & 2.86908 & 2.87053 \\ 
24 & 1.90402 & 2.81827 & 2.86908 & 2.87053 \\

\end{tabular}
\end{table}

\begin{table}
\caption{
Binding energies in the case of the MTVa potential
 for a system of three bosons.}
\label{tmt5}
\begin{tabular}{ccccc}

&  \multicolumn{4}{c}{\mbox{Angular momentum channels} }  \\
& 1 & 2 & 3 & 4 \\
\hline 
\mbox{This work} & 8.04251  & 8.22953 & 8.24978 & 8.25215 \\
Faddeev \cite{la1} & 8.0424 & 8.228   & 8.249   & 8.251 \\ 
\hline 
Faddeev \cite{gr1} & \multicolumn{4}{c}{ 8.25273}  \\
ATMS \cite{atms} & \multicolumn{4}{c}{ 8.26(1)}  \\
GFMC \cite{gfmc} & \multicolumn{4}{c}{ 8.26(1)}  \\
IDEA \cite{idea} & \multicolumn{4}{c}{ 8.25}  \\
SVM \cite{svm} & \multicolumn{4}{c}{ 8.2527}  \\
\end{tabular}
\end{table} 

\begin{table}
\caption{
Binding energies and root-mean-square radii 
for the MTVb potential for a system of three bosons.}
\label{tmt5b}
\begin{tabular}{cccccc}

& &  \multicolumn{4}{c}{\mbox{Angular momentum channels} }  \\
& & 1 & 2 & 3 & 4 \\
\hline 
$-E_B$ &\mbox{This work}    & 7.5398 & 7.7147   & 7.7338   & 7.7361 \\ 
     & Faddeev \cite{la1} & 7.540 & 7.714   & 7.733   & 7.735 \\ 
\hline 
$<r^2>^{1/2}$ &\mbox{This work}  & 1.7265 & 1.7117  & 1.7098   & 1.7095 \\ 
     & Faddeev \cite{la1} & 1.727 &  1.711  & 1.710   & 1.710 \\ 
\end{tabular}
\end{table}

\begin{table}
\caption{
Binding energies and root-mean-square radii
in the case of the Ali-Bodmer and  
Ali-Bodmer plus Coulomb potentials (Eqs.
(\ref{alibodmer}) and (\ref{alibodmerC}), respectively) 
for a system of three bosons.}
\label{tabcomp}
\begin{tabular}{cccc}

\mbox{Ali-Bodmer} & & \mbox{Without $v^C$} &  \mbox{With $v^C$} \\
\hline 
&\mbox{This work} & 5.181  & 2.871 \\
$-E_B$ & ATMS \cite{atms} &  5.18  &  \\
& SVM \cite{svm} &  5.18  & 2.872 \\
\hline
&\mbox{This work} & 2.434  & 2.517 \\
$<r^2>^{1/2}$ & ATMS \cite{atms} &  2.43  &  \\
& SVM \cite{svm} &  2.43  & 2.517 \\ 
\end{tabular}
\end{table} 

\begin{figure}
\caption{Contour $C$ for the integral in Eq. (\ref{contourint})
in case of the three-body bound-state problem. To the right of the 
dotted vertical line lies the (continuous) spectrum of 
$h^0_{\eta_\alpha}$, to its left the (discrete and continuous) 
spectrum of $h_{\xi_\alpha}$.}
\label{fig1}
\end{figure}

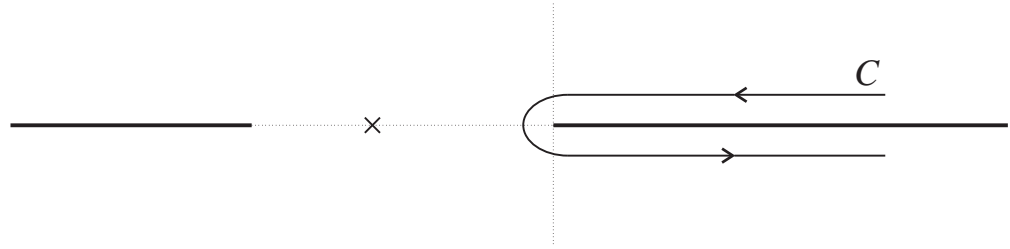
\end{document}